# Latent Moment Models for Recurrent Binary Outcomes: A Bayesian and Quasi-Distributional Approach

**Niloofar Ramezani** (ramezanin2@vcu.edu)
Department of Biostatistics, Virginia Commonwealth University

**Lori P. Selby** (lpselby@asu.edu)
School of Mathematics and Statistics, Arizona State University

**Pascal Nitiema** (pnitiema@asu.edu)
Department of Information Systems, Arizona State University

**Jeffrey R. Wilson** (jeffrey.wilson@asu.edu)
Department of Economics, Arizona State University

**Abstract**

Recurrent binary outcomes within individuals, such as hospital readmissions, often reflect latent risk processes that evolve over time. Conventional methods like generalized linear mixed models and generalized estimating equations estimate average risk but fail to capture temporal changes in variability, asymmetry, and tail behavior. We introduce two statistical frameworks that model each binary event as the outcome of a thresholded value drawn from a time-varying latent distribution defined by its location, scale, skewness, and kurtosis. Rather than treating these four quantities as nonparametric moment estimators, we model them as interpretable latent moments within a flexible latent distributional family. The first, BLaS-Recurrent, is a Bayesian model using the sinh–arcsinh distribution (a parametric family that provides explicit control over asymmetry and tail weight) to estimate latent moment trajectories; the second, QuaD-Recurrent, is a quasi-distributional approach that maps simulated moment vectors to event probabilities using a flexible nonparametric surface. Both models support time-dependent covariates, serial correlation, and multiple membership structures. Simulation studies show improved calibration, interpretability, and robustness over standard models. Applied to ICU readmission data from the MIMIC-IV database, both approaches uncover clinically meaningful patterns in latent risk, such as right-skewed escalation and widening dispersion, that are missed by traditional methods. These models provide interpretable, distribution-sensitive tools for longitudinal binary outcomes in healthcare while explicitly acknowledging that latent "moments" summarize—but do not uniquely determine—the underlying distribution.

## 1. Introduction

Hospital readmissions, treatment relapses, and adverse clinical events are examples of recurrent binary outcomes that are commonly studied in medical research. These events often reflect latent patient-level risk processes that evolve dynamically in response to physiological changes, treatment effects, and healthcare system exposures. Accurately modeling such risk is essential for effective triage, post-discharge planning, and intervention targeting.

Generalized linear mixed models (GLMMs) and generalized estimating equations (GEEs) are widely used for analyzing longitudinal binary data. GLMMs provide subject-specific inference through random effects, while GEEs offer marginal estimates based on a working correlation structure [1,2]. However, both approaches typically focus on modeling the mean outcome and assume a fixed or symmetric latent risk distribution. As a result, they may miss critical features such as increasing variability, distributional skewness, or tail intensification, which can signal elevated clinical risk.

Recent advances in distributional regression have expanded modeling beyond the mean to include variance, skewness, and kurtosis, enabling more nuanced understanding of outcome distributions [3,4]. While this has been applied successfully to continuous data, few methods extend it to binary or recurrent binary settings. Related models from psychometrics—such as item response theory (IRT) and dynamic factor models—allow for evolving latent traits but are often limited to unidimensional and symmetric latent structures [5], making them less suitable for irregular, clustered data typical in medical studies.

This paper introduces two complementary modeling frameworks for recurrent binary outcomes based on evolving latent risk distributions. The first, BLaS-Recurrent, is a fully Bayesian model that uses the sinh–arcsinh distribution (a flexible four-parameter family that provides interpretable control over location, scale, skewness, and tail weight) to estimate time-varying latent moments [6]. These parameters are treated as covariate-dependent components of a parametric latent distribution, rather than as nonparametric moment estimators, acknowledging that the first four moments alone do not uniquely determine an arbitrary distribution. The second, QuaD-Recurrent, is a quasi-distributional approach that calibrates event probabilities from simulated latent moment vectors via a nonparametric regression surface. Both models support time-varying covariates, serial correlation, and complex clustering structures including multiple membership effects [7]. We model each latent moments parameter using its own regression equation, following distributional-regression frameworks that allow covariates to influence location, scale, skewness, and tail weight.

Through simulation studies and application to ICU readmission data from the MIMIC-IV database [8], we show that these models outperform traditional approaches in calibration, interpretability, and robustness. They offer a flexible and clinically meaningful framework for understanding how risk evolves—not just in magnitude, but in shape—over time.

## 2. Background and Related Work

Modeling recurrent binary outcomes is central to medical research involving recurrent events such as hospital readmissions, symptom relapses, or repeated treatment failures. Two widely used approaches—GEEs and GLMMs—have been applied extensively to these settings. GEEs estimate marginal probabilities while accounting for within-subject dependence via a working correlation matrix, making them suitable for population-level inference [1]. GLMMs, in contrast, incorporate subject-specific random effects to model heterogeneity and within-subject dependence, yielding conditional inferences at the individual level [2].

Despite their widespread use, both frameworks are generally restricted to modeling mean-level risk and assume fixed-shape latent distributions, typically normal or logistic. This limits their ability to reflect dynamic risk variation or to capture higher-order features such as distributional skewness or tail intensification, which are often present in clinical populations undergoing change over time.

In continuous time, recurrent event modeling has evolved through extensions of survival analysis, including the Andersen–Gill model, the Prentice–Williams–Peterson model, and shared frailty models [3]. These approaches are valuable for handling censoring and time-varying covariates but focus on hazard rates rather than the shape of underlying risk distributions. As such, they provide limited insight into asymmetry or escalation dynamics that may characterize recurrence processes. Additional extensions in survival modeling, such as frailty-based Cox models, offer improved individual-level inference but remain limited in capturing latent distributional shape [3].

In the social sciences, latent trait models such as IRT and dynamic factor models offer a probabilistic foundation for modeling binary outcomes as reflections of latent states that evolve over time [4]. However, these models typically assume unidimensional, symmetric latent variables and are not well suited to irregular follow-up schedules or hierarchical structures commonly found in medical studies.

Recent advances in distributional regression have enabled regression modeling not only for the mean, but also for variance, skewness, and kurtosis of continuous outcomes [5,6]. While these methods have improved understanding of heteroskedasticity and tail behavior, their extension to thresholded binary outcomes remains limited—particularly in longitudinal or clustered contexts. Adaptations of GAMLSS for binary outcomes have begun to address these gaps, but are not yet widely applied in medical longitudinal settings [9].

The current work extends this modeling tradition by embedding each binary outcome in a time-varying latent distribution with flexible shape. Specifically, we estimate latent location, scale, skewness, and kurtosis as functions of time-dependent covariates, serial correlation, and nested or overlapping random effects [7]. The BLaS-Recurrent model assumes a parametric sinh–arcsinh latent structure [10], which provides interpretable control over asymmetry and tail weight and ensures that all admissible combinations of latent moments correspond to valid probability distributions. The QuaD-Recurrent model calibrates event probabilities via simulation, treating the four latent "moments" as covariate-dependent shape summaries rather than as nonparametric estimators, and explicitly acknowledging that they do not uniquely determine an arbitrary distribution. Both approaches support clinically interpretable, shape-sensitive modeling of dynamic risk.

## 3. Methods

We propose a latent variable framework for modeling recurrent binary outcomes, where each binary observation is viewed as the outcome of a thresholded draw from an evolving, subject-specific latent distribution. Unlike conventional models that focus solely on mean-level risk, our approach estimates time-varying location, scale, skewness, and tail-weight to describe the full shape of the underlying latent risk process. These quantities are treated as covariate-dependent latent moments within a parametric latent distributional family, rather than as nonparametric moment estimators, acknowledging that the first four moments alone do not uniquely determine an arbitrary distribution.

### 3.1 Latent Outcome Model

Let $Y_{it} \in \{0,1\}$ denote the binary outcome for individual $i = 1, \ldots, N$ at time $t = 1, \ldots, T$. We assume that each $Y_{it}$ arises as

$$Y_{it} = \mathbb{I}(Z_{it} > 0),$$

where $\mathbb{I}(.)$ denotes a binary-valued indicator function (equal to 1 when $Z_{it} > 0$ and 0 otherwise). The latent variable $Z_{it}$ follows a continuous distribution parameterized by four time-varying latent moments:

$$Z_{it} \sim f(z|\mu_{it}, \sigma_{it}, \nu_{it}, \tau_{it})$$

where $\mu_{it}$: latent location (mean), $\sigma_{it}$: latent scale (standard deviation), $\nu_{it}$: skewness parameter, $\tau_{it}$: kurtosis or tail weight parameter. These quantities are not raw empirical moments; rather, they are interpretable latent moments within the chosen latent distributional family. We explicitly acknowledge that the first four moments alone do not uniquely determine an arbitrary distribution, and infinitely many distributions may share the same moment quadruple. Our modeling approach therefore treats $(\mu,\sigma,\nu,\tau)$ as parameters of a parametric family—rather than as nonparametric moment estimators—ensuring that all admissible combinations correspond to valid probability distributions.

Each latent moment, $m_{it}$ for individual $i$ observed at time $t$, is modeled as a function of covariates and random effects:

$$m_{it} = \mathbf{x}_{it}^{(m)\prime} \boldsymbol{\beta}^{(m)} + b_i^{(m)} + u_{it}^{(m)}, \quad m \in \{\mu, \sigma, \nu, \tau\}$$

where $\mathbf{x}_{it}^{(m)\prime}$: covariate vector for moment $m$, $\boldsymbol{\beta}^{(m)}$: fixed effects, $b_i^{(m)}$: subject-level random intercept, $u_{it}^{(m)}$ are optional autoregressive error terms to accommodate serial dependence. To ensure positivity, the scale and tail-weight parameters are modeled using log-links,

$$\sigma_{it} = \exp(\tilde{\sigma}_{it}), \tau_{it} = \exp(\tilde{\tau}_{it}).$$

If included, $u_{it}^{(m)}$ follows a Gaussian AR(1) process with parameter $\rho_m$.

To ensure reproducibility and clarify identifiability, we explicitly define all notation and dimensions used in the latent regression structure. Let $p_m$ denote the number of covariates included in the regression for moment $m \in \{\mu, \sigma, \nu, \tau\}$. Then:

- $\boldsymbol{x}_{it}^{(m)}$ is a $p_m \times 1$ covariate vector
- $\boldsymbol{\beta}^{(m)}$ is a $p_m \times 1$ coefficient vector
- $b_i^{(m)}$ is a scalar random intercept for moment $m$
- $u_{it}^{(m)}$ is a scalar AR(1) error term
- Each individual contributes four random intercepts $\{b_i^{(\mu)}, b_i^{(\sigma)}, b_i^{(\nu)}, b_i^{(\tau)}\}$
- Total number of regression parameters is $\sum_m p_m$
- Total number of random effect parameters is $4N$

Although this yields a large number of latent random effects, identifiability is preserved because:
1. Each moment equation is conditionally independent given covariates and random effects.
2. The SAS family provides a one-to-one mapping between $(\mu, \sigma, \nu, \tau)$ and the latent distribution.
3. Random effects enter additively and are regularized through hierarchical priors.

**Admissibility constraints.** For arbitrary distributions with finite moments, the inequality kurtosis > skewness$^2$ + 1 must hold. Because we work within the SAS family, admissibility is automatically enforced: all combinations of $(\mu, \sigma, \nu, \tau)$ correspond to valid distributions without requiring explicit constraints. This is a key advantage of using a parametric latent family rather than modeling raw moments directly. Table 1 summarizes the roles, link functions, and interpretations of the four latent distribution parameters.

We adopt linear predictors for each shape parameter because:
- they provide interpretable covariate effects on each aspect of the latent distribution;
- positivity constraints for $\sigma$ and $\tau$ are enforced through log-links;
- the SAS transformation embeds $\nu$ and $\tau$ in a smooth, monotone mapping, making linear predictors appropriate for these shape parameters;
- this structure parallels distributional regression frameworks (e.g., GAMLSS), where each distributional parameter has its own regression equation.

**Table 1.** Latent distribution parameters and regression links

| Moment | Parameter | Link | Interpretation |
|---|---|---|---|
| Location | $\mu$ | Identity | Latent mean |
| Scale | $\sigma$ | Log | Latent SD (volatility) |
| Skewness | $\nu$ | Identity | Asymmetry of latent risk |
| Tail weight | $\tau$ | Log | Kurtosis / tail heaviness |

## 3.2 BLaS-Recurrent: Parametric Bayesian Model

In the BLaS-Recurrent model, we assume the latent variable follows the sinh–arcsinh (SAS) distribution [10], a flexible four-parameter family that provides interpretable control over location, scale, skewness, and tail-weight:

$$Z_{it} \sim SAS(\mu_{it}, \sigma_{it}, \nu_{it}, \tau_{it})$$

The SAS distribution is defined by the transformation

$$Z = \mu + \sigma \sinh\left(\frac{\operatorname{arcsinh}(Y) + \nu}{\tau}\right),$$

where $Y \sim N(0,1)$. This yields a unimodal distribution on $\mathbb{R}$ whose parameters have the following roles: location ($\mu$), scale ($\sigma$), skewness ($\nu$), and tail-weight ($\tau$). The corresponding CDF is

$$F_{\text{SAS}}(z \mid \mu, \sigma, \nu, \tau) = \Phi\left(\tau \operatorname{arcsinh}\left(\frac{z - \mu}{\sigma}\right) - \nu\right),$$

where $\Phi(\cdot)$ is the standard normal CDF.

The probability of event occurrence is given by:

$$Pr(Y_{it} = 1 \mid \mu_{it}, \sigma_{it}, \nu_{it}, \tau_{it}) = 1 - F_{SAS}(0 \mid \mu_{it}, \sigma_{it}, \nu_{it}, \tau_{it})$$

where $F_{SAS}$ denotes the cumulative distribution function of the SAS distribution.

The full likelihood across all subjects and time points is:

$$\mathcal{L} = \prod_{i=1}^{N} \prod_{t=1}^{T} [1 - F_{SAS}(0 \mid \mu_{it}, \sigma_{it}, \nu_{it}, \tau_{it})]^{Y_{it}} [F_{SAS}(0 \mid \mu_{it}, \sigma_{it}, \nu_{it}, \tau_{it})]^{1 - Y_{it}}$$

We place normal priors on all regression coefficients and weakly informative priors on variance components. For each moment $m \in \{\mu, \sigma, \nu, \tau\}$, let $\psi_m^2$ denote the variance of the subject-specific random intercepts $b_i^{(m)}$. We assign a Half-Cauchy(0, 2.5) prior to this variance parameter:

$$\beta^{(m)} \sim \mathcal{N}(0, \lambda_m^2 I) \text{ and } \psi_m^2 \sim \text{Half} - \text{Cauchy}(0, 2.5)$$

The hyperparameter $\lambda_m$ controls the degree of shrinkage for moment-specific covariate effects. The model is implemented in the Stan modeling language [11], and inference is conducted via Hamiltonian Monte Carlo using the Stan platform, enabling full posterior estimation of covariate effects, latent moment trajectories, and outcome probabilities [12].

The SAS family is adopted because it provides explicit and interpretable control over skewness and tail-weight, allowing the latent distribution to flexibly capture asymmetry and heavy-tailed behavior when present. The distribution reduces to the normal when $\nu = 0$ and $\tau = 1$, ensuring that the model nests the standard symmetric latent-variable formulation as a special case. Importantly, the SAS parameterization guarantees that all combinations of $(\mu, \sigma, \nu, \tau)$ correspond to valid probability distributions, thereby avoiding moment-constraint violations such as kurtosis $>$ skewness$^2 + 1$ that arise when modeling raw moments directly. The SAS distribution is unimodal with full support on $\mathbb{R}$, making it particularly suitable for threshold-based latent-variable models in which the entire real line must be represented without imposing artificial truncation or symmetry.

Model adequacy under the SAS specification can be evaluated using several diagnostic tools. Posterior predictive checks comparing observed and replicated binary outcomes provide a direct assessment of calibration. QQ-plots of posterior draws of $Z_{it}$ against fitted SAS quantiles allow visual evaluation of whether the latent distribution captures asymmetry and tail behavior. Moment-based summaries of the posterior latent distributions offer additional insight into whether the estimated latent moments align with empirical patterns. Finally, comparing SAS-based predictions with those obtained from the quasi-distributional model offers a complementary robustness check, as discrepancies may indicate latent-distributional features not fully captured by the parametric SAS family. Together, these diagnostics help determine whether the SAS distribution provides an adequate representation of the underlying latent risk process.

### 3.3 QuaD-Recurrent: Quasi-Distributional Model

The QuaD-Recurrent model does not impose a parametric form for the latent distribution. Instead, it uses simulation to approximate the mapping from latent moment vectors to binary event probabilities:

$$Pr(Y_{it} = 1 | \mu_{it}, \sigma_{it}, \nu_{it}, \tau_{it}) \approx \hat{\Phi}(\mu_{it}, \sigma_{it}, \nu_{it}, \tau_{it})$$

The approximation surface $\hat{\Phi}$ is constructed offline by generating a large set of latent-parameter combinations and simulating the corresponding event probabilities. Specifically, we

1. Generate a grid of moment combinations ($\mu$, $\sigma$, $\nu$, $\tau$) spanning clinically plausible ranges;
2. Simulate latent variables Z ~ f (z | $\mu$, $\sigma$, $\nu$, $\tau$) (where $f$ is the SAS density used for internal consistency) using the same SAS family to ensure internal consistency, and compute Pr(Z>0) for each grid point;
3. These simulated probabilities are then smoothed using an interpolation method such as penalized splines or Gaussian process regression to obtain a continuous approximation surface.

This surface is used in a pseudo-likelihood:

$$\mathcal{L}_{pseudo} = \prod_{i=1}^{N}\prod_{t=1}^{T} \widehat{\Phi}[\boldsymbol{\theta}_{it}]^{Y_{it}}\left[1 - \widehat{\Phi}[\boldsymbol{\theta}_{it}]\right]^{1-Y_{it}}$$

Where $\theta_{it} = (\mu_{it}, \sigma_{it}, \nu_{it}, \tau_{it})$ is the latent moment parameter vector for individual $i$ at time $t$. Posterior sampling is conducted using standard Bayesian methods. QuaD-Recurrent approximates the mapping from latent moment vectors to event probabilities, rather than specifying a full likelihood. Although the model is not likelihood-based in the conventional sense, this approach offers strong robustness to latent distribution misspecification and supports flexible inference [13].

### 3.4 Random Effects and Multiple Membership
To accommodate complex data structures—such as patients exposed to multiple hospital units—we include multiple membership random effects [14,15]. For latent moment $m$, the subject-specific random effect is:

$$b_i^{(m)} = \sum_{g=1}^{G} w_{ig}\gamma_g^{(m)}, \quad \gamma_g^{(m)} \sim \mathcal{N}(0, \theta_m^2), \quad \sum_{g=1}^{G} w_{ig} = 1$$

where $w_{ig}$ represents the proportion of time individual $i$ was exposed to group $g$. This multiple-membership formulation allows each subject's latent trajectory to reflect contributions from several overlapping groups, weighted by their exposure patterns. This structure integrates seamlessly into the hierarchical model and preserves interpretability of latent moment trajectories across heterogeneous care environments.

### 3.5 Summary
These two modeling frameworks provide flexible, shape-sensitive, and interpretable methods for analyzing recurrent binary outcomes in settings where risk profiles evolve over time. The BLaS-Recurrent model supports full likelihood inference with distributional interpretability, while the QuaD-Recurrent model offers computational efficiency and robustness to latent-distribution misspecification. Both are well-suited to medical applications where shape dynamics—such as escalation, asymmetry, or long-tail recurrence—convey prognostic value and clinical urgency.

### 4. Simulation Study
We conducted a simulation study to evaluate the performance of the BLaS-Recurrent and QuaD-Recurrent models in estimating time-varying latent risk structures and predicting recurrent binary outcomes. The study focused on four core objectives: 1. latent moment recovery — assessing how accurately each model estimates the evolving location, scale, skewness, and kurtosis of the latent risk distribution; 2. Predictive calibration — evaluating how well predicted event probabilities align with observed outcomes; 3. Robustness to latent distribution misspecification — testing sensitivity to deviations from the assumed or approximated latent shape; and 4. Scalability — examining model stability and runtime performance across varying sample sizes. These design elements reflect practical challenges frequently encountered in applied longitudinal data analysis, particularly where dynamic risk patterns inform clinical decisions.

## 4.1 Data Generation

Binary outcomes were generated for samples N=250,500, and 1000 individuals, each observed at T=6 time points. At each time point, the binary outcome $Y_{it}$ was defined as a thresholded latent variable:

$$Y_{it} = \mathbb{I}(Z_{it} > 0), \; Z_{it} \sim SAS(\mu_{it}, \sigma_{it}, \nu_{it}, \tau_{it})$$

The covariate structure included a static predictor $x_i \sim \mathcal{U}(-1,1)$ and a time-varying predictor $x_{it} \sim \mathcal{U}(-1,1)$. Each latent moment was modeled as a linear function of these covariates and a subject-specific random effect. For instance, the location was defined as

$$\mu_{it} = \beta_0^{(\mu)} + \beta_1^{(\mu)} x_i + \beta_2^{(\mu)} x_{it} + b_i^{(\mu)}$$

Analogous structures were used for scale, skewness, and kurtosis. Scale and kurtosis components were transformed using exponentiation to ensure positivity. The random effects $b_i^{(\mu)}$ were independently drawn from normal distributions with moderate variance, reflecting realistic between-subject variability.

This setup allowed us to evaluate the models' ability to recover latent moment dynamics and generate well-calibrated predictions under controlled yet clinically plausible conditions.

## 4.2 Misspecification Scenarios

To assess robustness, we introduced two latent distribution misspecification scenarios. In the first, $Z_{it}$ followed a skew-t distribution with dynamic skew and heavy tails. In the second, $Z_{it}$ was drawn from a Gaussian mixture distribution with time-varying component weights. While the covariate and latent moment structures remained unchanged, the latent distribution violated the parametric assumptions underlying the BLaS-Recurrent model. This setup allowed us to evaluate model calibration and adaptability under model misfit—conditions commonly encountered in real-world clinical settings.

## 4.3 Competing Models

We benchmarked BLaS-Recurrent and QuaD-Recurrent against three widely used alternatives:
- A generalized linear mixed model (GLMM) with a logistic link and random intercepts
- A generalized estimating equation (GEE) approach with exchangeable working correlation
- A decision tree classifier with time-aware splitting, serving as a nonparametric baseline

All Bayesian models were estimated using Hamiltonian Monte Carlo with four chains and 2,000 iterations (1,000 warm-up). The QuaD-Recurrent model used a simulation-trained probability surface based on 10,000 moment vectors and corresponding binary probabilities, fit via penalized splines.

## 4.4 Evaluation Metrics

Performance was evaluated along four dimensions. Predictive accuracy was assessed using the area under the ROC curve (AUC), Brier score, and calibration slope and intercept, averaged across time points. Latent moment recovery was evaluated by calculating bias, root mean squared error (RMSE), and 95% credible interval coverage for each parameter. Robustness was tested under misspecified latent distributions by comparing predictive metrics and coverage. Finally, scalability was assessed by examining model performance across increasing sample sizes with a fixed follow-up length. All metrics were averaged over 200 replications. Posterior summaries were aggregated across individuals and time points to evaluate inference stability and variability.

### 4.5 Summary
This simulation design provided a comprehensive evaluation of each model's ability to recover latent moment trajectories and generate well-calibrated predictions under both ideal and challenging conditions. The results serve as a benchmark for interpreting performance in the real-world clinical dataset analyzed in the following section.

## 5. Results
We summarize findings from the simulation study across 200 replications, comparing BLaS-Recurrent and QuaD-Recurrent with standard approaches in terms of predictive performance, latent moment recovery, robustness to misspecification, and scalability. Unless otherwise noted, results are reported for N=500 and T=6. Additional simulations at N = 250 and N = 1000 showed consistent qualitative patterns in predictive performance and moment recovery. Detailed tables are provided in the Supplementary Material.

### 5.1 Predictive Accuracy and Calibration
Table 2 reports average AUC, Brier score, and calibration diagnostics. Both BLaS-Recurrent and QuaD-Recurrent outperformed the GLMM, GEE, and decision tree baselines. BLaS-Recurrent yielded the strongest overall performance, with an AUC of 0.79, Brier score of 0.169, and a calibration slope of 0.97. QuaD-Recurrent followed closely with an AUC of 0.78 and slope of 0.94. GLMM and GEE showed lower discrimination and under-calibration, while the decision tree exhibited poor calibration despite a modest AUC.

**Table 2.** Predictive performance for N=500, T=6
(*Averaged across 200 replications*)

| Model | AUC | Brier Score | Calibration Slope | Calibration Intercept |
|---|---|---|---|---|
| GEE | 0.73 | 0.191 | 0.81 | −0.04 |
| GLMM | 0.75 | 0.186 | 0.87 | −0.03 |
| Decision Tree | 0.77 | 0.195 | 0.66 | −0.06 |
| QuaD-Recurrent | 0.78 | 0.172 | 0.94 | −0.01 |
| BLaS-Recurrent | 0.79 | 0.169 | 0.97 | 0.00 |

### 5.2 Latent Moment Recovery
Table 3 summarizes bias and 95% posterior interval coverage for each latent moment. Both models recovered location and scale with high accuracy. BLaS-Recurrent showed superior performance on skewness and kurtosis, with coverage rates above 89%, while QuaD-Recurrent produced slightly wider intervals and lower coverage for these higher-order parameters.

### 5.3 Covariate Effect Estimation
Both models accurately recovered fixed effects for each latent moment. Posterior means were centered around the true values, and 95% credible intervals achieved nominal coverage. BLaS-Recurrent consistently yielded narrower intervals, particularly for skewness and kurtosis, due to its full likelihood formulation. QuaD-Recurrent performed comparably but exhibited more conservative uncertainty estimates, particularly for higher-order parameters.

**Table 3.** Bias and 95% coverage for latent moments (N=500)

| Moment | Model | Bias | Coverage (%) |
|---|---|---|---|
| Mean ($\mu$) | BLaS-Recurrent | 0.02 | 94.8 |
|  | QuaD-Recurrent | 0.03 | 93.7 |
| Variance ($\sigma^2$) | BLaS-Recurrent | 0.05 | 92.1 |
|  | QuaD-Recurrent | 0.06 | 91.5 |
| Skewness ($\nu$) | BLaS-Recurrent | 0.07 | 90.2 |
|  | QuaD-Recurrent | 0.09 | 88.6 |
| Kurtosis ($\tau$) | BLaS-Recurrent | 0.10 | 89.3 |
|  | QuaD-Recurrent | 0.13 | 86.9 |

### 5.4 Robustness to Latent Misspecification

Under skew-t and Gaussian mixture distributions, QuaD-Recurrent preserved strong calibration and discrimination, maintaining an AUC of 0.78 and a calibration slope of 0.93. BLaS-Recurrent showed a slight drop in AUC (0.77) and reduced credible interval coverage for higher-order parameters, but still outperformed GLMM and GEE. Both standard methods substantially underestimated tail probabilities and showed degraded calibration (slopes < 0.80), confirming their sensitivity to latent distributional misspecification.

### 5.5 Sample Size Sensitivity

With N=250, both models remained stable in prediction metrics, though uncertainty in skewness and kurtosis increased modestly. At N=1000, posterior precision improved across all parameters. BLaS-Recurrent showed sharper interval contraction, while QuaD-Recurrent scaled more efficiently due to its pre-trained simulation surface. These patterns are consistent with the broader simulation results, which showed qualitatively similar behavior across N = 250, 500, and 1000. Overall, both models demonstrated robustness to sample size variation.

### 5.6 Sensitivity to Model Specification

We examined two simplified variants of BLaS-Recurrent: one excluding skewness ("NoSkew"), and another excluding both skewness and kurtosis ("NoTail"). Under symmetric data-generating processes, predictive performance was comparable. However, when skewness or excess kurtosis was present, the full model outperformed reduced variants. It achieved a 0.03 higher AUC and better-calibrated risk estimates, particularly in the upper tails, highlighting the value of capturing higher-order structure in latent risk distributions.

### 5.7 Summary

Both BLaS-Recurrent and QuaD-Recurrent delivered strong and complementary performance. BLaS-Recurrent offered precise latent moment recovery and sharp posterior inference, while QuaD-Recurrent demonstrated robustness to misfit and superior scalability. Together, these models address key limitations of traditional methods and offer practical tools for modeling evolving binary risk in medical settings.

## 6. Real-World Application: ICU Readmissions in MIMIC-IV

To demonstrate the clinical utility of the proposed models, we applied BLaS-Recurrent and QuaD-Recurrent to data from the MIMIC-IV database [8], focusing on predicting repeated 30-day ICU readmissions. This application illustrates how distribution-aware models can uncover dynamic risk patterns in high-stakes healthcare settings.

### 6.1 Cohort and Outcome Definition

We constructed a retrospective cohort of N=1,218 adult patients (aged ≥ 18) with at least three post-discharge observation windows. The final dataset comprised 5,631 patient-time observations. The binary outcome was defined as an unplanned ICU readmission within each 30-day window.

### 6.2 Covariates

The analysis included both static and time-varying covariates. Static variables included age, sex, and insurance type. Time-varying variables, aggregated within each 30-day window, included maximum white blood cell (WBC) count, maximum creatinine, average heart rate, and average systolic blood pressure. An additional covariate captured the proportion of the prior hospital stay spent in the ICU. A continuous time index (days since initial ICU discharge) was used to model temporal progression. All continuous variables were standardized before model fitting.

### 6.3 Model Estimation

Both BLaS-Recurrent and QuaD-Recurrent were fit using the full covariate set to estimate time-varying latent location ($\mu$), scale ($\sigma$), skewness ($\nu$), and kurtosis ($\tau$). Random intercepts were included to account for within-subject dependence. Multiple-membership random effects were incorporated using a weighted structure to represent patients' exposure to different hospital units [14,15].

BLaS-Recurrent was estimated via Hamiltonian Monte Carlo using Stan (four chains, 2,000 iterations). QuaD-Recurrent used a pre-trained simulation surface $\hat{\Phi}(\mu, \sigma, \nu, \tau)$ constructed from 20,000 latent moment-parameter vectors and their associated binary probabilities. For baseline comparison, a GLMM with a logistic link and random intercepts was also estimated.

### 6.4 Predictive Performance

Table 4 presents 10-fold cross-validated performance metrics. Both latent moment models outperformed the GLMM. BLaS-Recurrent achieved the highest calibration slope (0.98) and lowest Brier score (0.169), while QuaD-Recurrent produced nearly identical AUC and better computational efficiency. The GLMM under-calibrated high-risk predictions and showed reduced discrimination.

**Table 4.** Predictive performance on held-out ICU readmission windows

| Model | AUC | Brier Score | Calibration Slope | Log Loss |
|---|---|---|---|---|
| GLMM | 0.72 | 0.192 | 0.84 | 0.583 |
| QuaD-Recurrent | 0.76 | 0.174 | 0.95 | 0.543 |
| BLaS-Recurrent | 0.77 | 0.169 | 0.98 | 0.534 |

### 6.5 Latent Risk Dynamics and Interpretability

Posterior estimates from BLaS-Recurrent revealed clear clinical patterns. Location ($\mu$) increased with age, creatinine, and ICU exposure—indicating elevated baseline risk. Scale ($\sigma$) grew over time, suggesting increased variability in patient trajectories. Skewness (ν) was associated with elevated WBC counts, producing right-skewed latent distributions and elevated event probabilities. Kurtosis (τ) decreased with extended ICU stays, yielding flatter, long-tailed risk profiles.

Figure 1 illustrates the evolution of the latent distribution for a representative patient across four follow-up windows. The curves show progressive rightward shift and dispersion, capturing subtle but clinically relevant risk dynamics that were previously not captured by standard models. Although the posterior mean trajectories of $Z_{it}$ in Figure 1 lie above zero, the posterior distributions retain substantial variability, and their credible intervals cross the threshold; as a result, event probabilities remain meaningfully below one, and the trajectories still convey clinically relevant changes in latent risk.

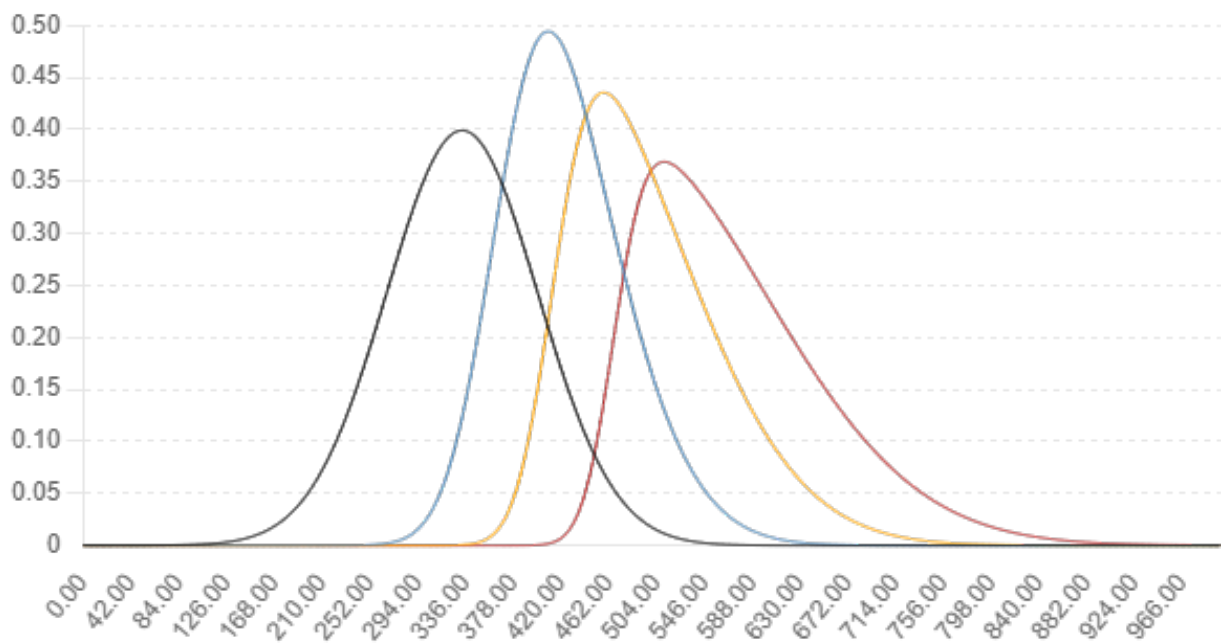

**Figure 1.** Posterior distributions of the latent risk variable $Z_{it}$ across four consecutive 30-day observation windows for a representative patient in the MIMIC-IV cohort. Each curve represents the estimated density from the BLaS-Recurrent model, incorporating covariates and patient-specific random effects. The dashed vertical line denotes the decision threshold at Z=0. Over time, the distribution shifts rightward and becomes more dispersed and right-skewed, indicating escalating and asymmetric readmission risk that is not captured by traditional models.

## 6.6 Clinical Implications

The proposed models provide interpretable outputs that can support clinical decision-making. Patients with moderate mean latent risk but pronounced skewness may benefit from enhanced monitoring, as their latent risk structure indicates potential for sudden deterioration. Similarly, increasing scale may signal growing physiological instability, while decreased kurtosis may reflect diffuse or poorly localized risk, warranting caution in discharge planning. By decomposing risk into its distributional components, these models offer actionable insight unavailable from mean-based predictions. Because each moment parameter maps to a distinct clinical intuition—e.g., volatility, asymmetry, or tail risk—these models may support transparent communication between data scientists and frontline clinicians. QuaD-Recurrent also offers advantages for deployment in time-sensitive or resource-constrained settings due to its efficiency and robustness to distributional assumptions.

## 6.7 Summary

Both BLaS-Recurrent and QuaD-Recurrent revealed clinically relevant, time-varying latent risk structures in ICU readmission data. They provided improved calibration and discrimination over standard models, and uncovered escalation and volatility patterns that may enhance early detection and personalized care planning. These results demonstrate the value of incorporating distributional moment dynamics in modeling recurrent outcomes in healthcare.

## 7. Discussion

Recurrent binary outcomes in healthcare—such as hospital readmissions or repeated treatment failures—often reflect risk processes that evolve over time and exhibit features beyond simple shifts in average probability. While GLMMs and GEEs are widely used for analyzing such outcomes, they focus on mean-level trends and assume fixed-shape latent structures. As a result, they may overlook clinically important patterns of volatility, asymmetry, and tail behavior.

This study introduces a modeling framework that extends beyond mean-based prediction by estimating time-varying latent location, scale, skewness, and kurtosis. The BLaS-Recurrent model, grounded in a Bayesian specification using the sinh–arcsinh distribution, supports full posterior inference for all four moment parameters. The alternative QuaD-Recurrent model takes a quasi-distributional approach, leveraging a calibrated simulation surface to estimate binary probabilities without strong distributional assumptions. Together, these methods provide interpretable tools for analyzing recurrent binary outcomes with evolving moment structure.

Simulation results showed that both models achieved superior calibration and discrimination compared to GLMMs and GEEs. BLaS-Recurrent offered sharper inference for higher-order moments, while QuaD-Recurrent maintained stable performance even under latent misspecification. These strengths were further validated in a real-world application to ICU readmission data, where both models uncovered risk dynamics—such as increasing skewness and scale—not captured by conventional models.

A key strength of these frameworks is interpretability. By decomposing predicted risk into components that describe not just magnitude but also distributional form, the models offer clinical insight into whether risk is concentrated, diffuse, or prone to escalation. This makes them suitable for decision support in settings where explainability and calibration are essential.

Several limitations merit consideration. First, estimation of skewness and kurtosis from thresholded binary outcomes can be sensitive to sample size and signal strength. While our simulations suggest that higher-order moments are recoverable when present, regularization or moment-selection strategies may be helpful in small or sparse datasets. Second, BLaS-Recurrent incurs computational cost due to full posterior sampling. Although feasible for moderate datasets, larger-scale applications may benefit from variational or amortized inference methods. QuaD-Recurrent, while not providing full posterior inference, offers a robust and scalable alternative in such settings.

Although we focused on binary outcomes, the modeling framework is general. Extensions to ordinal, count, and time-to-event outcomes are possible by adapting the threshold mechanism or linking function. Multivariate formulations and models incorporating feedback or competing risks could also support more comprehensive health trajectory modeling.

In summary, BLaS-Recurrent and QuaD-Recurrent provide a principled, interpretable framework for modeling recurrent binary outcomes with dynamic latent risk. By capturing not just the level but the moment structure of evolving risk, these models support calibrated and clinically actionable prediction in longitudinal settings. Future work may explore integration with electronic health record systems for real-time risk monitoring.

**Supplementary Material**

**Supplementary Table S1.** Predictive performance for N = 250 and N = 1000 for baseline models (oracle, GEE, decision tree) Averaged across 200 replications. These simulations assess sample-size sensitivity and complement the main results reported for N = 500 in Table 2.

| N | Model | AUC | Brier | Calibration Slope | Calibration Intercept |
|---|---|---|---|---|---|
| 250 | oracle | 0.8145 | 0.1761 | 0.2145 | 0.5005 |
| 250 | GEE | 0.7279 | 0.2102 | 0.2180 | 0.5001 |
| 250 | Decision Tree | 0.7767 | 0.1897 | 0.0368 | 0.4933 |
| 1000 | oracle | 0.8180 | 0.1747 | 0.2138 | 0.5019 |
| 1000 | GEE | 0.7318 | 0.2094 | 0.2175 | 0.5001 |
| 1000 | Decision Tree | 0.7468 | 0.2031 | 0.0968 | 0.4988 |